\documentclass[12pt,preprint]{aastex}

\begin{document}

\shorttitle{Polarization Variability in 0420$-$014}
\shortauthors{D'Arcangelo et al.}

\title{Rapid Multiwaveband Polarization Variability in the Quasar PKS 0420$-$014: Optical Emission
from the Compact Radio Jet}

\author{Francesca D. D'Arcangelo\altaffilmark{1}, Alan P. Marscher\altaffilmark{1},
Svetlana G. Jorstad\altaffilmark{1,2}, Paul S. Smith\altaffilmark{3}, Valeri M. Larionov\altaffilmark{2},
Vladimir A. Hagen-Thorn\altaffilmark{2}, Eugenia N. Kopatskaya\altaffilmark{2},
G. Grant Williams\altaffilmark{4},
Walter K. Gear\altaffilmark{5}}

\altaffiltext{1}{Institute for Astrophysical Research, Boston University, Boston, MA 02215;
fdarcang@bu.edu, jorstad@bu.edu, marscher@bu.edu}
\altaffiltext{2}{Sobolev Astronomical Institute, St. Petersburg State University, Universitetskij pr. 28,
198504 St. Petersburg, Russia; vml@vl1104.spb.edu, hth@vg3823.spb.edu, kopac@astro.spbu.ru}
\altaffiltext{3}{Steward Observatory, The University of Arizona, Tucson, AZ 85721-0065; psmith@as.arizona.edu}
\altaffiltext{4}{Multiple Mirror Telescope Observatory, The University of Arizona, Tucson, AZ 85721-0065; 
gwilliams@mmto.org}
\altaffiltext{5}{School of Physics and Astronomy, Cardiff University, 5 The Parade, Cardiff CF2 3YB,
Wales, UK; Walter.Gear@astro.cf.ac.uk}

\begin{abstract}
An 11-day monitoring campaign in late 2005 reveals clear correlation in polarization between the
optical emission and the region of the intensity peak (the ``pseudocore'')
at the upstream end of the jet in 43 GHz Very Long Baseline Array images
in the highly variable quasar PKS 0420$-$014. The electric-vector position angle (EVPA) of the pseudocore
rotated by about $80^\circ$ in four
VLBA observations over a period of nine days, matching the trend of the optical EVPA. In addition,
the 43 GHz EVPAs agree well with the optical values when we correct the former for Faraday rotation. 
Fluctuations in the polarization at both wavebands are consistent with the variable emission arising from
a standing conical shock wave that compresses magnetically turbulent plasma
in the ambient jet. The volume of the variable component is the same at both wavebands, although
only $\sim 20$\% of the total 43 GHz emission arises from this site. The remainder of
the 43 GHz flux density must originate in a separate region with very low polarization.
If 0420$-$014 is a typical case, the nonthermal optical emission from blazars originates primarily
in and near the pseudocore rather than closer to the central engine where the flow collimates
and accelerates.
\end{abstract}


\section{Introduction}
Our knowledge of the jets of active galactic nuclei (AGN) becomes less certain as we consider regions closer
to the central engine. However, these regions are the sites of especially
interesting physical phenomena, such as
acceleration and collimation of the jet flow, formation of the observed ``core'' of the jet,
energization of relativistic electrons, and changes in the geometry of the magnetic field. While it is
widely accepted that synchrotron emission is the main mechanism behind continuum production, establishing
the location of this emission at different wavebands has been problematic.
Imaging with very long baseline interferometry (VLBI) can resolve different components of the jet at radio 
wavelengths. In addition, variability of the flux density and spectral shape provides information on
the small-scale behavior at all wavebands, but the inability to produce high quality images at
submilliarcsecond resolution at wavelengths shorter than 3 mm allows
multiple interpretations that are compatible with the data.

There is, however, an emerging method for
establishing the location of the emission at short wavelengths. Similarities of unique polarization
signatures of features in both VLBI images and optical measurements provide a possible connection
between the two wavebands.  Recent work by \citet{gab06} and \citet{jor07}, that extend earlier
studies by \citet{war84}, \citet{sit85}, \citet{gab94}, and \citet{lis00}, reveals a correlation
between VLBI pseudocore (see below) and optical polarization angle in a number of blazars.
These results suggest that the two emission regions are either cospatial or separate but with both
threaded by magnetic fields having similar geometries. [Here we follow \citet{jones88},
in referring to the compact feature at the upstream end of the jet in VLBI images as the ``pseudocore''. This
may be an actual stationary feature, such as a standing shock wave \citep[see, e.g.,][]{dm88,caw06}, or
the point in the diverging jet flow where the optical depth is of order unity \citep{bk79,kon81}.]

The most direct way to establish cospatiality of the emission at different wavebands is to observe
coordinated variability of polarization that is seen both in the VLBI images and in measurements at
shorter wavelengths. In this {\it Letter} we report observations of polarization variations in both
the optical emission and the pseudocore as measured on 43 GHz Very Long Baseline Array
(VLBA) images of the quasar PKS~0420$-$014 ($z=0.915$; \citeauthor{hb93} \citeyear{hb93}) that span
nine days, with no significant time lag. These results place the
nonthermal optical emission within the 7 mm pseudocore of the jet.

\section{Observations}

We collected optical data at two sites. At the Steward Observatory 1.55 m Kuiper telescope, we
used the SPOL spectropolarimeter \citep{sss92}, making a total of nine measurements from 2005 October
23 to November 3. A 600 lines/mm grating covered an entire diffraction order of $\sim 400$-800 nm
at a resolution of 1.7 nm. The spectropolarimetry used slit widths of either $2''$ or $3''$ along
with a $10''$ spectral extraction aperture. Differential spectrophotometry relative to field stars
calibrated by \citet{sb98} employed a $5.1''\times12''$ (or $14''$) aperture. Details of the reduction
procedure can be found in, e.g., \citet{sshf03}.
We also performed polarimetry and photometry at $R$ band on the AZT-8 70 cm telescope at the Crimean
Astrophysical Observatory. The signal-to-noise ratio was low, however, owing to the faintness of
0420$-$014 ($R$ = 17.8-18.3) during the campaign. We therefore do not display the data here, but
note that they are consistent with the Steward Observatory results.

We observed 0420$-$014 at 43 GHz with the VLBA at four epochs during the campaign: October 24,
28 and November 1, 2. After correlation at the Array Operations Center of the National Radio Astronomy
Observatory (NRAO) in Socorro, New Mexico, we
followed the procedures described in \citet{j05} to create and analyze the resultant images.  We refer the
polarization position angle measurements to a stable feature in the VLBA polarized intensity image
of the quasar CTA102.  We find good agreement between the adjusted angles of several objects with
archival data and with contemporaneous Very Large Array measurements
(available on website http://www.vla.nrao.edu/astro/calib/polar/). In our analysis,
we add $31\degr$ to the polarization angle in the pseudocore of 0420$-$014 to compensate
for Faraday rotation, based on the estimate of rotation measure by \citet{jor07}.

We also observed 0420$-$014 at two epochs (October 24 and 27) with the Heterodyne Spectrometer
at the James Clerk Maxwell Telescope (JCMT) at a frequency of 230 GHz. The sensitivity of the system
was sufficient only to determine the flux density of 0420$-$014 relative to the planets Uranus and Mars.

\section{Analysis}

\begin{figure}
\plotone{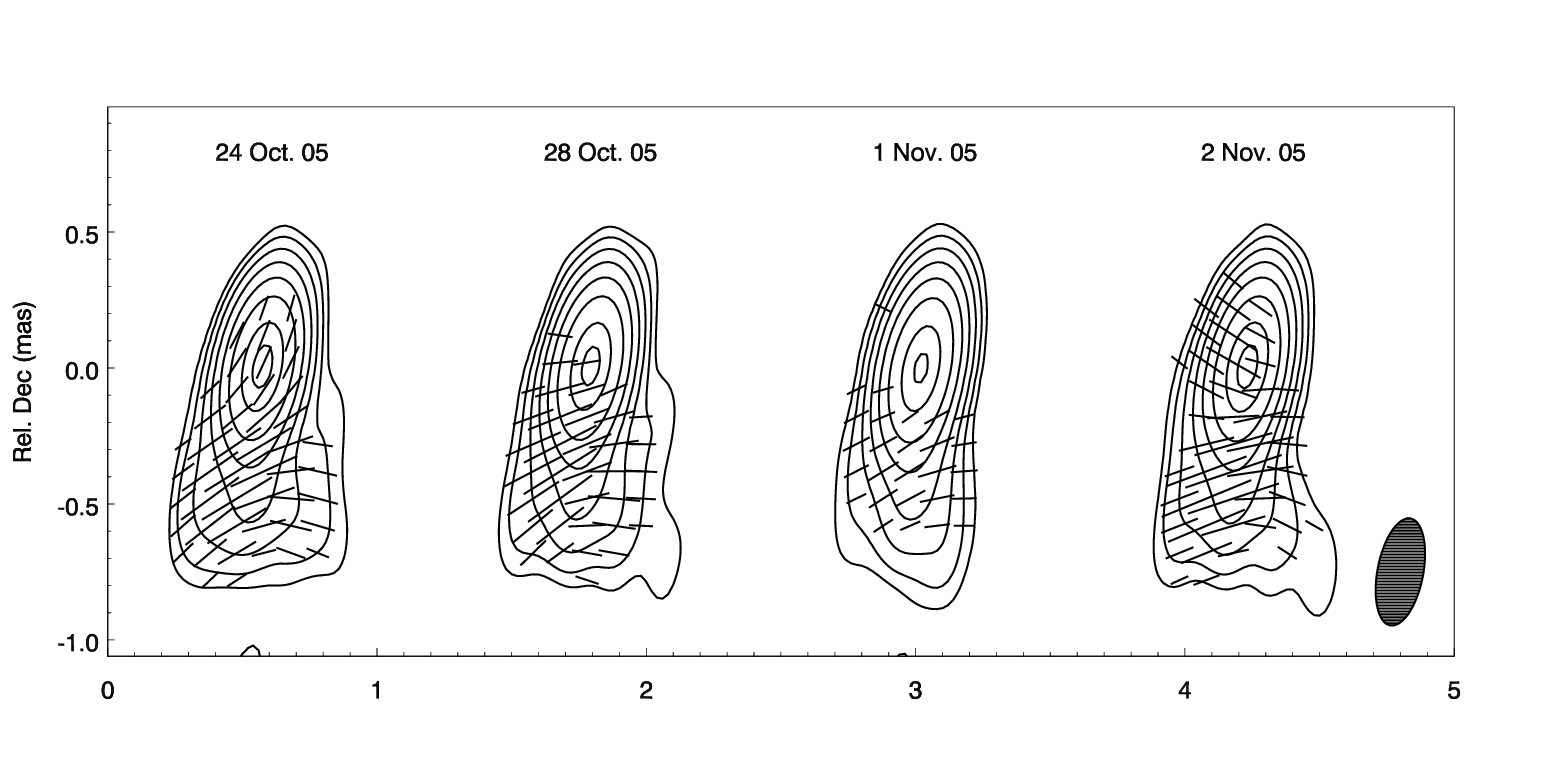}
\caption{43 GHz (7 mm) images of 0420$-$014 at the four epochs of VLBA observations. We refer to
the region surrounding the peak in total intensity as the ``pseudocore.'' The sacle of the
horizontal, right ascension axis, are the same as for the vertical, declination axis. Contours
represent 1, 2, 4, 8, 16, 32, 64, and 90\% of the peak total intensity
of 2.0 Jy/beam, while the sticks indicate polarized intensity (length of 0.1 mas corresponds to
6.2 mJy/beam) and direction of electric vector.  Correction for statistical bias and for
Faraday rotation toward the pseudocore of $-31\degr$ at 43 GHz has not been applied to the displayed
polarization vectors. As noted in the text, the very low polarization of the pseudocore in the
2005 November 1 image is an artifact of a pronounced change in polarization position angle during
the ten-hour observation.
The rms polarization noise level is 1.8 mJy/beam. Dimensions of the restoring beam (shown as a shaded
ellipse in the lower right corner) are 0.40$\times$0.17 mas along position angle $-11\degr$. 
\label{fig1}}
\end{figure}

\begin{figure}
\plotone{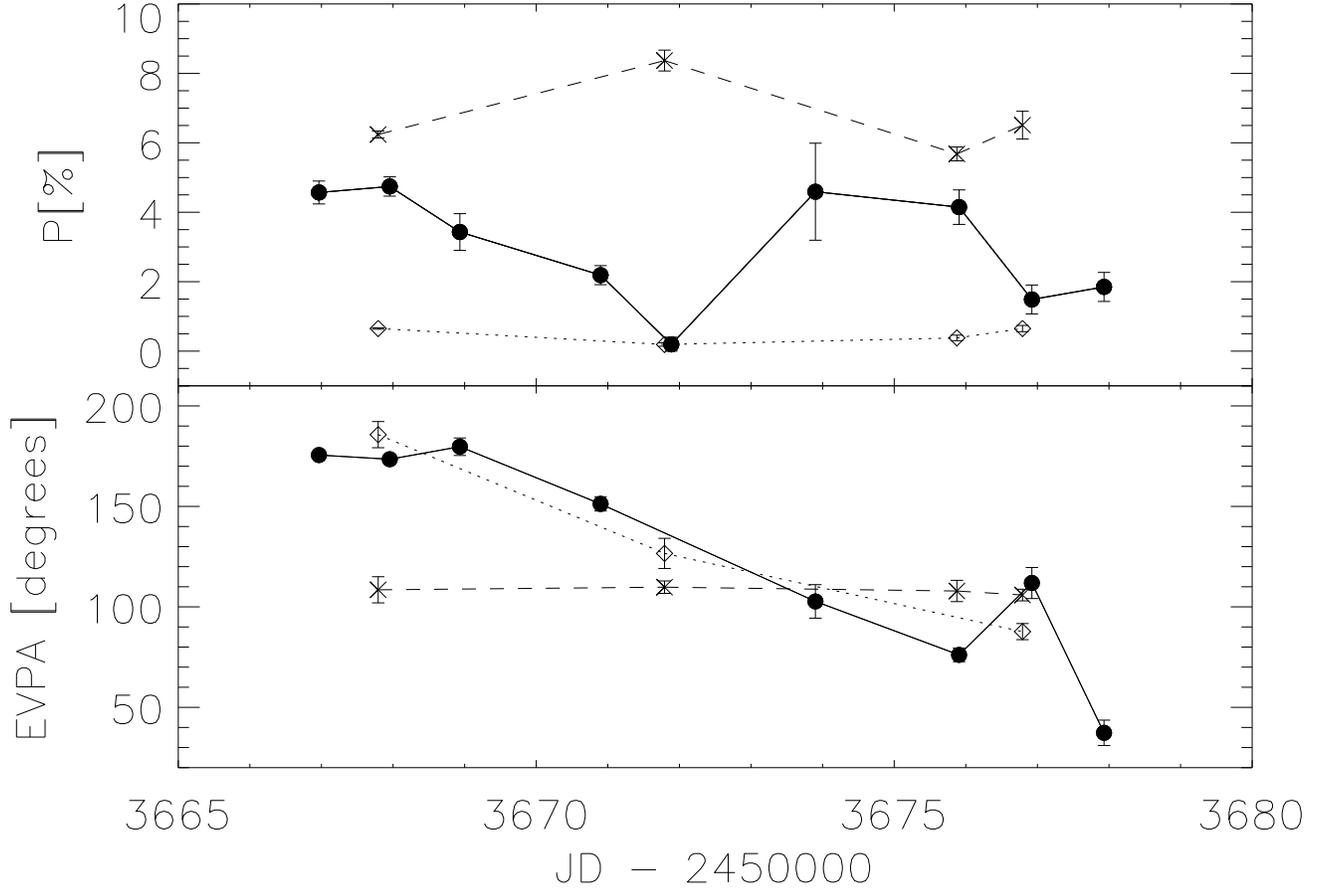}
\caption{{\it Top:} Variation of percent polarization with time of 0420$-$014 in the optical band
(450-750 nm; filled circles and solid line), in the 7 mm pseudocore (open diamonds and dotted line),
and in the first jet component in the 7 mm image ($\times$ and dashed line). The
degrees of polarization are corrected for statistical bias \citep{war74}.
{\it Bottom:} Variation of EVPA with time of the same three components of PKS 0420$-$014.
The EVPA is not plotted when $P/\sigma(P) < 1.5$ or for the pseudocore at JD 2453675.9 when
it varied greatly during the VLBA observation.
A correction of $31\degr$ has been added to the EVPAs of the pseudocore
to compensate for Faraday rotation.  For reference, JD 2453670 is 2005 October 26.
\label{fig2}}
\end{figure}

On milliarcsecond (mas) scales, the 43 GHz structure of 0420$-$014 consists of a compact pseudocore and
a jet with knots that propagate away from the pseudocore on paths that eventually curve southward
\citep{j05}. The VLBA images from our campaign (Fig. \ref{fig1}) feature a broad jet extending to
0.7 mas south of the pseudocore. The pseudocore, with a flux density of $\sim 2$ Jy, has a low degree of
polarization, $P < 1\%$, but a highly variable EVPA (see Fig. \ref{fig2}).  The first distinct jet
component, a 0.45 Jy knot lying 0.4 mas south of the pseudocore, is more highly polarized
($P \sim 7\%$). Its degree and angle of
polarization remained steady throughout the campaign to within the uncertainties
(cf. Fig. \ref{fig2}), providing a check on the calibration. The very low polarization of the pseudocore
at the third epoch of VLBA data (Julian date 2453675.9) is an artifact of the long observation time
($\sim 10$ hours) required to obtain a high-quality image. Upon splitting the observation in two,
we find that the EVPA varied by $\sim 90\degr$ from the first half to the second half of the observation,
thereby cancelling the polarization in the resultant image shown in Figure \ref{fig1}. The degree
of polarization of the pseudocore displayed in Figure \ref{fig2} for this epoch is the average between
the two halves. For the remaining VLBA images, the EVPA did not change by more than $2\sigma$ between
the first and second half of the observation.

The optical polarization underwent pronounced variations in both $P$ and EVPA, as shown
in Figure \ref{fig2}. Of particular note is
that the 43 GHz pseudocore and optical EVPAs varied in a similar manner, following each other remarkably
well over a range of $\sim 80\degr$. By eye, we see that a straight line through the three radio points
(the polarization of the pseudocore was too weak to measure on November 1) provides a reasonable
representation of the overall optical trend. Upon performing a least-squares linear fit to the EVPA
{vs.} time plots that takes
into account the uncertainties in the data, we find that the pseudocore polarization vector rotates by
$-10.5\degr\pm 0.8\degr$ day$^{-1}$, while the optical vector rotates by $-11.1\degr\pm 0.3\degr$ day$^{-1}$.
There is no significant time lag between the 43 GHz and optical polarization variations.
Despite the similarity in EVPA at the two wavebands, the degree of polarization $P$ is considerably higher
($\langle P \rangle=3.0\pm0.2$\%) at the optical band than in the 7 mm pseudocore
($\langle P \rangle=0.47\pm0.03$\%).  This implies that $\sim 80\%$ of the 43 GHz emission from the
pseudocore arises from a separate, essentially unpolarized region that is blended with the pseudocore at
the angular resolution of the VLBA. We suggest that this region is optically thick with a tangled magnetic
field, in which case the polarization would be very close to zero. A higher resolution 86 GHz polarized
intensity image should reveal this component.

\section{Discussion}

The striking correspondence between variations in polarization at the two wavebands locates the primary
site of optical nonthermal emission in this quasar within the 7 mm pseudocore. As emphasized by
\citet{mar06}, this generally lies well downstream of the central engine and ``true core,'' since the
self-absorption turnover frequency of the overall synchrotron spectrum of 0420$-$014 occurs at a shorter
wavelength, $\sim 3$ mm \citep{bloom94}. Our flux density measurements of 2.5 Jy at 7 mm
and 1.1 Jy at 1.35 mm during the campaign indeed indicate that the turnover was at a wavelength longer
than 1.35 mm. The true core might be the end of the bulk acceleration and collimation
zone \citep[see][]{vk04} beyond which the jet is conical with essentially constant mean flow Lorentz factor
\citep{mar80,jor07}.

Magnetic turbulence is the most straightforward way to explain the low fractional polarization generally
observed in the compact jets of quasars at radio wavelengths \citep[e.g.,][]{jones88,hug05}.
We approximate the turbulence as $N$ cells of equal volume, each with uniform but randomly oriented magnetic
field. The randomness dilutes most of the polarization of the
source such that the mean degree of polarization $\langle{P}\rangle = P_{\rm max}N^{-1/2}$, where
$P_{\rm max} = (\alpha+1)/(\alpha+5/3)$ is the maximum linear polarization of incoherent synchrotron radiation
and $\alpha$ is the spectral index of the total flux density spectrum, $F_\nu \propto \nu^{-\alpha}$
\citep{Burn66,jones88,hug91}.
The standard deviation about the mean value $\sigma(P) = 2^{-1/2}\langle{P}\rangle$. The mean
fractional optical polarization of $\sim 3\%$ then implies that $\sim 600$ cells participate in the
synchrotron radiation at this waveband.

The polarization of the pseudocore could vary if there were a changeover in turbulent cells involved in
the emission during the course of our campaign. This can be accomplished if the radiating electrons
are energized by a shock wave that passes through the turbulence \citep{jones85,jones88,MGT92} or vice
versa. However, only in the unlikely case of a face-on propagating shock wave would the partial ordering
of the magnetic field parallel to the shock front not be apparent from the observer's perspective;
otherwise, the compression would affect the polarization by establishing a preferred EVPA
\citep[e.g.,][]{hug91}. Furthermore, electrons accelerated at a shock front
lose energy as they advect away from the
front, with those that attain higher energies---and therefore emit at higher frequencies---having shorter
radiative lifetimes. This should cause the optical emission to occupy a much
smaller volume ($\propto \lambda^{1/2}$) than does the mm-wave emission \citep{mar85}, so that the latter
should involve $\sim 60,000$ cells, or $\sim 100$ times as many as at optical wavelengths.
Although this is roughly compatible with the lower mean polarization at 7 mm, it cannot explain the
similarity in EVPA in the two wavebands, since the additional factor of $\sim 100$ cells should
produce a mean polarization angle at 7 mm that is essentially independent of the optical EVPA.

In order for a shock plus turbulence model to be valid, the thickness of the section of shocked plasma
imposed by gas dynamics \citep[the shocked region ends at a rarefaction;
see, e.g.,][]{dm88,gom95} must be comparable to the radiative lifetime of the
optically emitting electrons times the speed at which the compressed plasma drifts from the acceleration
zone at the shock front.  The timescale of variability of the optical polarization
$t_{\rm var} \approx 4$ days (see Fig. \ref{fig2}). If we adopt the
apparent speed and angle to the line of sight of the jet in 0420$-$014 derived from VLBA
monitoring by \citet{j05}, $\beta_{\rm app} \approx 11$ and $\theta \approx 3\degr$ (Doppler factor
of 16 and flow Lorentz factor $\Gamma \approx 11$), a
section of jet plasma of thickness $x \sim 0.8$ pc passes by a stationary shock during a time
interval of 4 days in the observer's reference frame. The synchrotron radiative lifetime of the
electrons at optical wavelengths $t_{\rm s} \approx 0.18(B$/0.1~G)$^{-3/2}$ yr, where we estimate the
magnetic field $B \sim 0.1$ G by
calculating the value corresponding to equipartition between the magnetic and relativistic electron energy
densities \citep[see][for evidence that the brightness temperatures of pseudocores of blazars are consistent
with equipartition]{h06}. The energized electrons travel a distance $\Gamma ct_{\rm s} \approx 0.6$ pc before
they can no longer radiate in the optical band (the $\Gamma$ factor accounts for Lorentz contraction of the
jet length in the plasma frame). This is essentially the same as the value of $x$ estimated above, so that
nearly all of the turbulent cells in the emission region are replaced every 4 days, as required by the
variability. The length of the section compressed by the shock can be less than the distance traversed by the
electrons radiating at optical wavelengths if the shock results from the body pinch-mode
instability owing to an oscillating pressure mismatch at the boundary between the jet and the external
medium \citep{dm88}. As seen in numerical gas dynamical simulations \citep{gom95}, such shocks have a
conical geometry if the jet is sufficiently close to being circularly symmetric. The partial ordering of
the magnetic field owing to compression by a conical shock contributes negligibly to the polarization for
very small viewing angles \citep{caw06} such as $3\degr$, the value derived for 0420$-$014 by \citet{j05}.

The apparent rotation of the polarization vector can be purely stochastic, as found by \citet{jones85},
who analyzed a model similar
to that described above. We have performed a number of simulations similar to those of
Jones et al., incorporating 600 cells, each with the same density and strength of a randomly oriented
magnetic field. The simulations produce artificial plots of percent polarization and EVPA vs. time.
Rotations by $100\degr$ or more occur $\sim 10\%$ of the time. Since 0420$-$014
was only one of 13 objects observed, the probability of catching one source at the start of a long
rotation episode is reasonably high if nearly all blazars behave in a similar manner.

Our observations could also be explained if a feature, e.g., a shock, with a partially ordered magnetic
field were to translate down a jet that twists by $\sim 140\degr$ over the length of $\sim 9$ pc that the
feature would move in 11 days \citep[see, e.g.,][]{gkw92}. This is not implausible, since curved jets are
common in 43 GHz images of blazars \citep[e.g.,][]{mar02,j05}. However, such a model does not naturally
explain the variations in degree of polarization and the fluctuations in EVPA about the
rotation trend (see Fig. \ref{fig2}).
A similar criticism argues against apparent rotation from changing aberration owing to acceleration of a
polarized knot as it propagates down the jet \citep{bk79}. Furthermore, monthly VLBA monitoring
throughout 2006 failed to detect any bright knot propagating down the jet that would correspond
to a shock wave moving through the pseudocore during our campaign. 
We can also eliminate a simple two-component model in which two nearly orthogonally polarized regions
vary in relative flux density such that the EVPA rotates from that of one feature to that of the other.
The maximum rotation possible in such a model is less than $90\degr$, while the optical EVPA changed by 
considerably more than $100\degr$ during the 11 nights of our campaign (see Fig. \ref{fig2}).

\section{Summary and Conclusions}
Our observation of concurrent polarization position angle at optical wavelengths and 7 mm as it swings by
$\sim 80\degr$ over an interval of nine days demonstrates that the main region producing
polarized optical emission is the same as that giving rise to the pseudocore at 43 GHz. During this
rotation, the degree of polarization varied and the EVPA fluctuated
about the mean trend in a manner consistent with a chaotic magnetic field. We propose a model that explains
the variations via changeovers in the $\sim 600$ turbulent cells involved in the emission over a period
of a few days. This is accomplished as
the plasma advects through a standing, conical shock where electrons are accelerated at the
front. The polarized emission at 43 GHz must originate
in the same volume as does the optical radiation, a situation that requires the region compressed
by the shock to be limited by gas dynamics rather than by radiative energy losses. This agrees with
gas dynamical simulations that show alternating regions of conical shocks and rarefactions in jets with
circular symmetry \citep{gom95}. We therefore conclude that the 43 GHz pseudocore in 0420$-$014 corresponds
to a standing shock system rather than the site where the optical depth is roughly unity. This is likely to
be the case in many blazars, since stationary features near the pseudocore in parsec-scale jets are common
\citep{j01}.

We are in the process of analyzing data from other objects observed during the campaign as well as objects
from a similar campaign in 2006 March-April.  These data will determine how
frequently rapid, correlated variations in optical and high-frequency radio EVPAs occur in
blazars. If this behavior is common, then we can search for similarity in polarization in a
number of blazars to identify the
locations of optical synchrotron emission on VLBI images. This would allow us to use correlations between
optical or mm-wave and X-ray as well as $\gamma$-ray variations in total flux density to determine the
location(s) of nonthermal emission across a factor of $\sim 10^{12}$ in frequency \citep{mj05,m06a}.

\acknowledgments
This material is based on work at Boston University supported by the National Science Foundation under grant 
AST-0406865, and at St. Petersburg State University supported by grant no. 05-02-17562 from the Russian
Fund for Basic Research. P.S.S. acknowledges support from NASA contract 1256424.
The VLBA is an instrument of the National Radio Astronomy Observatory, a facility of the National Science
Foundation operated under cooperative agreement by Associated Universities, Inc. The JCMT is operated by
the Joint Astronomy Centre on behalf of the Particle Physics and Astronomy Research Council of the United
Kingdom, the Netherlands Organization for Scientific Research, and the National Research Council of Canada.

\clearpage

\clearpage

\end{document}